\documentclass[twocolumn,prl,amsmath,amssymb,superscriptaddress,showpacs]{revtex4}

\usepackage{graphicx}
\usepackage{dcolumn}
\usepackage{bm}

\newcommand {\vB}{{\bf B}}
\newcommand {\vH}{{\bf H}}
\newcommand {\vE}{{\bf E}}
\newcommand {\vD}{{\bf D}}

\begin{document}

\title{Negative refraction from quasi-planar chiral inclusions}

\author{R. Marqu\'es}
\email{marques@us.es} %
\affiliation{Dept.~of Electronics and Electromagnetism. University
of Sevilla, 41012-Sevilla (Spain)}
\author{L. Jelinek}
\affiliation{Dept.~of Electromagnetic Field. Czech Tech.
University, Prague (Czech Rep.)}
\author{F. Mesa}
\affiliation{Dept.~of Applied Physics 1. University of Sevilla,
41012-Sevilla (Spain)}

\date{\today}

\begin{abstract}

This letter proposes a quasi-planar chiral resonator suitable for
the design of negative refractive index matamaterial. It is presented an
analytical model for the determination of its polarizabilities, and the
viability of negative refraction in chiral and racemic arrangements
with the proposed inclusions is analyzed. The present analysis is expected
to pave the way to the design of negative refractive index
matamaterials made of a single kind of inclusions feasible from
standard photo-etching techniques.

\end{abstract}


\maketitle

The main aim of this letter is to explore the possibility of
obtaining negative refraction from a random arrangement of
quasi-planar chiral inclusions. Artificial bi-isotropic chiral
media made of random arrangements of metallic chiral inclusions
are known for long, after the former works of K.~Lindmann
\cite{Lindmann}. More recently in \cite{Tretyakov} balanced (or
\emph{racemic}) mixtures of such type of inclusions were proposed as
a way to obtain negative refractive index metamaterials. The
general conditions for negative refraction of plane waves at the
interface between ordinary and chiral media were analyzed in
\cite{Mackay}, and the focusing of circularly polarized light by a
chiral slab was studied in \cite{Monzon}. The main advantage of
chiral elements in order to provide negative refraction is that
only one kind of inclusions is necessary to obtain negative values
of $\epsilon$ and $\mu$. An additional advantage would come from
the application of conventional printed circuit fabrication
techniques to manufacture such inclusions. For such purpose, a
quasi-planar design would be desirable.

\begin{figure}[tbph]
\centering
\includegraphics[width=60mm]{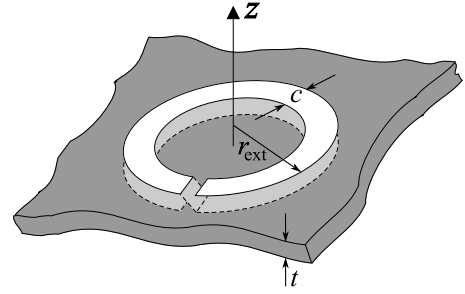}
\caption{The proposed inclusion is formed by two identical
conducting rings, etched on both sides of a dielectric substrate,
and connected by a via in order to obtain an helicoidal shape.}
\label{inclusion}
\end{figure}

The proposed inclusion is shown in Fig.~\ref{inclusion}. It is the
broadside-coupled version of the two turns spiral resonator (2-SR)
previously proposed by some of the authors as a metamaterial
element \cite{Baena}. The analysis in that paper shows that the
proposed element can be characterized by a quasi-static $LC$
circuit, where $L$ is the inductance of a single ring with the
same radius and width as the inclusion, and $C=2\pi r
C_{\text{pul}}$ is the total capacitance between the rings.
However, there are two main differences between the structure of
Fig.~\ref{inclusion} and the 2-SR analyzed in \cite{Baena}. First,
due to the broadside coupling, the distributed capacitance between
the rings can be made very large, which will reduce the electrical
size of the inclusion near the resonance. Second, when the element
is excited near the resonance, in addition to a strong magnetic
dipole there also appears a strong electric dipole oriented
parallel to the former one. This latter property comes from the
strong electric field between the upper and lower rings that
appears near the resonance.

Neglecting losses, and following the analysis in \cite{Baena}, the
circuit equation for the total current in the element (i.e., for
the sum of the currents excited on both rings, which must be
angle-independent \cite{Baena}) is given by
\begin{equation}\label{I-equation}
      \left(\frac{1}{j\omega C} + j\omega L \right)I=\Sigma\;,
\end{equation}
where $\Sigma$ stands for the external voltage excitation. For magnetic
excitation: $\Sigma = -j\omega\pi r^2 B_z^\text{ext}$, where
$r$ is the mean radius of the inclusion. For electric excitation:
$\Sigma = t\,C_0/C \,E_z^\text{ext}$, where $t$ is the substrate
thickness and $C_0$ is the total capacitance between the rings in
the absence of the dielectric substrate \cite{nota}. From such
equations, the following electric and magnetic moments excited in the
inclusion when it is submitted to some external electric and/or
magnetic fields can be obtained:
\begin{eqnarray}
   m_z &=&\alpha_{zz}^\text{mm}B_z^\text{ext} -
    \alpha_{zz}^\text{em}E_z^\text{ext} \nonumber
  \\
  \label{m-p}
   p_z &=&\alpha_{zz}^\text{ee} E_z^\text{ext} +
   \alpha_{zz}^\text{em}B_z^\text{ext} \;,
\end{eqnarray}
where
\begin{eqnarray}
     \label{alphamm}
     \alpha_{zz}^\text{mm} &=& \frac{\pi^2r^4}{L}
    \left(\frac{\omega_0^2}{\omega^2}-1\right)^{-1}
      \\
      \label{alphaem}
     \alpha_{zz}^\text{em} &=& \pm j\pi
r^2tC_0\frac{\omega_0^2}{\omega}
     \left(\frac{\omega_0^2}{\omega}-1\right)^{-1}
      \\
     \label{alphaee}
     \alpha_{zz}^\text{ee} &=& t^2 C_0^2L \frac{\omega_0^4}{\omega^2}
      \left(\frac{\omega_0^2}{\omega^2}-1\right)^{-1} \,,
\end{eqnarray}
with $\omega_0=\sqrt{1/LC}$ being the frequency of resonance. From
(\ref{alphamm})--(\ref{alphaee}) follows that
\begin{equation}\label{alpha-prop}
\alpha_{zz}^\text{mm}\alpha_{zz}^\text{ee} =
        -\left(\alpha_{zz}^\text{em}\right)^2 \;,
\end{equation}
which will be useful in the following \cite{nota2}. When $N$
chiral inclusions are assembled in a random way, the resulting
medium becomes bi-isotropic with constitutive relations given by
\begin{eqnarray}
      \vD &=& \varepsilon_0\varepsilon_r\vE + j
\sqrt{\varepsilon_0\mu_0}\;\kappa \vH
\;;\;\;\;\;\;\; \varepsilon_r = (1+\chi_e) \label{D} \\
\vB &=& -j \sqrt{\varepsilon_0\mu_0} \; \kappa \vE + \mu_0\mu_r\vH
\;;\; \mu_r=(1+\chi_m) \label{B}\,.
\end{eqnarray}
The electric, $\chi_e$, magnetic, $\chi_m$, and cross, $\kappa$,
susceptibilities are related to the inclusion polarizabilities through
\begin{equation} \label{chi-kappa}
   \chi_e = \frac{N}{\Delta
\varepsilon_0}\frac{\alpha_{zz}^\text{ee}}{3} \;;\;\;\; \chi_m =
\frac{N\mu_0}{\Delta}\frac{\alpha_{zz}^\text{mm}}{3} \;;\;\;\;
\kappa = \pm j \frac{N}{\Delta} \sqrt{\frac{\mu_0}{\varepsilon_0}}
\frac{\alpha_{zz}^\text{em}}{3} \,,
\end{equation}
where the factor $1/3$ arises from the random arrangement, and
$\Delta$ is a common factor that depends on the homogenization
procedure. From (\ref{alpha-prop}) and (\ref{chi-kappa}) follows
that
\begin{equation}\label{chi-property}
\chi_e(\omega)\chi_m(\omega) = [\kappa(\omega)]^2\;.
\end{equation}

\begin{figure}[tbph]
\centering
\includegraphics[width=50mm]{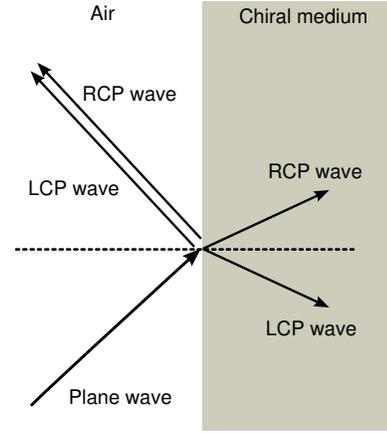}
\caption{\label{NR} Illustration of the negative refraction of a
linearly polarized wave at the interface with a chiral
metamaterial made of inclusions as that shown in
Fig.~\ref{inclusion}. Only one of the two eigenwaves that can
propagate in the chiral medium shows negative refraction, and the
reflected wave is elliptically polarized.}
\end{figure}

As is well known, the general dispersion equation for plane
waves in lossless chiral media is
\begin{equation}\label{k}
 k = \pm
k_0\left(\sqrt{(1+\chi_e)(1+\chi_m)}\pm\kappa\right)\,,
\end{equation}
where $k_0=\omega\sqrt{\epsilon_0\mu_0}$. The four solutions of
(\ref{k}) correspond to right- and left-hand circularly polarized
waves, depending on the sign of $\kappa$. In order to avoid
complex solutions of (\ref{k}), and therefore forbidden frequency
bands for plane wave propagation, it would be desirable that
$\chi_e(\omega) = \chi_m(\omega)$. According to (\ref{chi-property}) this
implies that
\begin{equation}\label{property-2}
    \chi_e = \chi_m  = |\kappa| \,.
 \end{equation}
The general condition for backward-wave propagation is found to be
\cite{Mackay}
\begin{equation} \label{condition}
  \sqrt{\epsilon_r\mu_r}\pm\kappa <0 \,,
\end{equation}
where the sign of the square root must be chosen negative if both
$\epsilon_r$ and $\mu_r$ are negative. According to
(\ref{condition}), if $\kappa^2 > |\epsilon_r\mu_r|$ only one of
solutions of (\ref{k}) can be a backward-wave and, therefore, will
experience negative refraction at the interface with an ordinary
media. This is indeed the case when (\ref{property-2}) is
satisfied and $\chi_e$, $\chi_m$, are both negative. In such case,
negative refraction will take place for only one of the eigenmodes
of (\ref{k}), provided that $\chi_e = \chi_m = |\kappa| < -0.5$.
This condition is less restrictive than the condition for ordinary
media (for instance, for a balanced mixture of inclusions of
opposite helicity), namely, $\chi_e,\,\chi_m\,< -1$. The price to
pay for this enlargement of the bandwidth is that only one of the
solutions of (\ref{k}) shows negative refraction. Such scenario is
illustrated in Fig.~\ref{NR}, where an incident linearly polarized
wave is considered.

\begin{figure}[tbph]
\centering
\includegraphics[width=77mm]{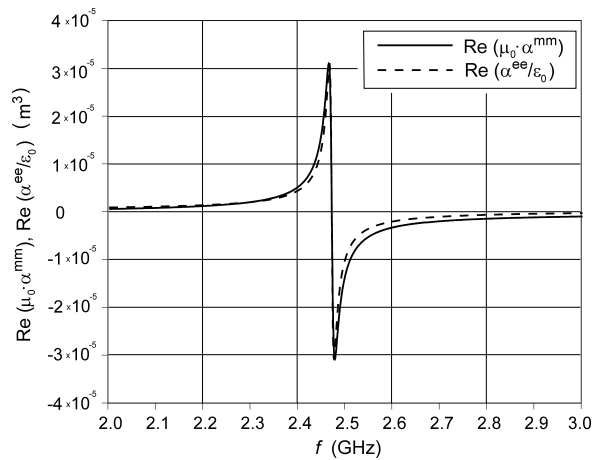}
\caption{\label{alphas} Numerical determination of
$\mu_0\alpha_{zz}^\text{mm}$ and
$\alpha_{zz}^\text{ee}/\epsilon_0$ for the inclusion shown in
Fig.\ref{inclusion} with the parameters given in the text.}
\end{figure}

Returning now to the inclusions, it is found from
(\ref{chi-kappa}) that condition (\ref{property-2}) is satisfied
provided that
\begin{equation}\label{property-3}
    c^2\alpha_{zz}^\text{ee}(\omega) =
    \alpha_{zz}^\text{mm}(\omega) = \pm jc\alpha_{zz}^\text{em}(\omega)\;,
\end{equation}
where $c$ is the velocity of light in vacuum. In priciple, this
condition is compatible with (\ref{alphamm})--(\ref{alphaee}).
Actually, we have tried to obtain a particular design satisfying
such condition by using the analytical expressions for $L$ and
$C_\text{pul}$ reported in \cite{Marques-APL}. A substrate with
permittivity similar to vacuum (a foam for instance) was chosen in
order to simplify computations. With this substrate
($\epsilon=\epsilon_0$) a suitable design is: width of the strips
$c=2\,$mm, external radius $r_\text{ext} = r + c/2 = 5\,$mm, and
separation between strips $t=2.35\,$mm. Following
\cite{Marques-APL}, the frequency of resonance of the proposed
configuration should be about $2.3\,$GHz. It gives an electrical
size of $\sim \lambda/13$ for the inclusion, which is acceptable
for a practical metamaterial design. In order to check our
analytical results, the electric and magnetic polarizabilities of
the inclusions have been numerically determined following the
procedure described in \cite{Lukas}. This procedure mainly
consists in placing the particle inside a TEM waveguide and to
compute the polarizabilities from the reflection and transmission
coefficients of the loaded waveguide (see \cite{Lukas} for more
details). The results for the meaningful quantities
$\mu_0\alpha_{zz}^\text{mm}$ and
$\alpha_{zz}^\text{ee}/\epsilon_0$ are shown in Fig.~\ref{alphas}.
These results clearly confirms the conclusions of our analytical
model. The cross polarizations cannot be numerically determined
following the method described in \cite{Lukas}. However, the
equality between the meaningful quantities
$\mu_0\alpha_{zz}^\text{mm}$ and
$|\sqrt{\mu_0/\epsilon_0}\,\alpha_{zz}^\text{em}|$ can be shown by
comparing the reflection coefficient for the co- and the
cross-polarized waves when the particle is placed inside a
metallic waveguide of square cross-section. If the waveguide is
wide enough, so as the wave impedance approaches that of free
space, and the particle is placed with its axis perpendicular to
the waveguide walls, the equality of both reflection coefficients
implies the equality of the above quantities. Numerical
calculations (not shown) made with the commercial electromagnetic
solver CST Microwave Studio confirms this prediction.

\begin{figure}[tbph]
\centering
\includegraphics[width=77mm]{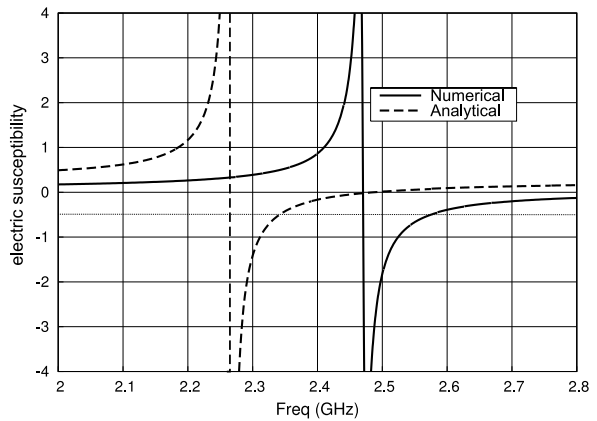}
\caption{\label{chi} Analytical and numerical results for the
electric susceptibility $\chi_e$ of a random arrangement of chiral
inclusions as those shown in Fig.~\ref{inclusion}. The parameters
of the inclusions are given in the text and are the same as in
Fig.~\ref{alphas}. The average volume per inclusion is
$V=12^{3}\,\text{mm}^3$.}
\end{figure}

In order to evaluate the frequency bandwidth for negative
refraction in a metamaterial made of a random arrangements of the
proposed inclusions, the electric susceptibility $\chi_e$ of such
medium has been computed from (\ref{chi-kappa}) with $\Delta=1$.
Although this approximation is rather rough, it is clear from the
general form of (\ref{chi-kappa}) that any other homogenization
procedure (for instance, a generalized Clausius-Mossotti one)
would give similar qualitative results. The dimensions and
characteristics of the inclusions are those previously reported,
and the number of inclusions per unit volume is
$N=(12)^{-3}\,\text{mm}^{-3}$. Both the analytical and the
numerical results obtained from the data of Fig.~\ref{alphas} are
shown in Fig.~\ref{chi}. From the analysis and the numerical
results reported in the previous paragraphs directly follows that
the curves (not shown) for the magnetic $\chi_m$ and the cross
susceptibility $\kappa$ must be quite similar. Although some
differences appear between the analytical and numerical results
shown in Fig.~\ref{chi}, its qualitative agreement is apparent. In
both cases a significant negative refraction frequency band
appears for both the random and the racemic mixtures. As it was
already mentioned, such frequency bands are limited by the
straight lines $\chi_e=-0.5$ and $\chi_e=-1$ respectively (see
Fig.\ref{chi}).

In summary, the feasibility of manufacturing negative refractive
index metamaterials from a random arrangement of chiral
quasi-planar inclusions has been analyzed. It has been proposed an specific
design with the advantage of being easily manufactured from
standard photo-etching techniques. Also it has been shown that such
design provides the necessary behavior for all the resonant
polarizabilities in order to produce a significant negative
refractive index bandwidth near the resonance.

\bigskip

\begin{acknowledgments}
This work has been supported by the Spanish Ministry of Education
and Science by project contract TEC2004--04249--C02--02.
\end{acknowledgments}


\begin{thebibliography}{12}

\bibitem{Lindmann} K.F.Lindmann. \emph{Annalen der Physik}, {\bf 63},
621 (1920).

\bibitem{Tretyakov} S.A.Tretyakov. \emph{Analytical modelling in
applied electromagnetics}, Artech House, Norwood MA (2003).

\bibitem{Mackay} T.G.Mackay. \emph{Microwave and Opt. Tech. Lett.},
{\bf 45}, 120 (2005).

\bibitem{Monzon} C.Monzon and D.W.Forester. \emph{Phys. Rev. Lett.},
{\bf 95}, 123904 (2005).

\bibitem{Baena} J.D. Baena, R. Marqu\'es, F. Medina, and J. Martel.
\emph{Phys. Rev. B}, {\bf 69}, 014402 (2004).

\bibitem{nota} The factor $C_0/C$ appears because, when a
parallel-plate capacitor is excited by a normal external field,
the elelectric field inside the capacitor is just the external one
multiplied by the above factor.

\bibitem{nota2} The inclusion also presents a non-resonant
electric polarizability in the transverse $z$-plane. Since this
polarizability is almost constant with frequency, and not very
large, it can be neglected in a first approximation.

\bibitem{Marques-APL} R. Marqu\'es, F. Mesa, J. Martel, and F. Medina.
\emph{IEEE Trans. Antennas and Propag.}, {\bf 51}, 2572 (2003).

\bibitem{Lukas} L.Jelinek, J.B.Baena, R.Marqu\'es, J.Zehentner
\emph{Proc. of the 36th European Microwave Conf.}, 983 (2006).

\end{thebibliography}
\end{document}